\documentstyle[12pt]{article}
\input{epsf}

\newcommand{\beq}{\begin{eqnarray}}
\newcommand{\eeq}{\end{eqnarray}}

\newcommand{\ket}{\rangle}
\newcommand{\bra}{\langle}
\newcommand{\tr}{\, {\rm tr}\, }

\begin{document}

\begin{center}
{\large Baryon Spectra
in Deformed Oscillator Quark Model\\ }
\vspace*{0.5cm}
 {A. Hosaka\footnote{e-mail address: hosaka@la.numazu-ct.ac.jp} \\
Numazu College of Technology, Numazu, Shizuoka 410, Japan}

\vspace*{0.5cm}
{H. Toki and M. Takayama\\
Research Center for Nuclear Physics (RCNP), Osaka University,
Ibaraki, Osaka 567, Japan}
\end{center}
\vspace*{1cm}
\abstract{We study theoretically the baryon spectra in terms of a
deformed oscillator quark (DOQ) model.
This model is motivated by
confinement of quarks and chiral symmetry breaking, which are the most
important non-perturbative phenomena of QCD.
The minimization of the DOQ
hamiltonian with respect to the deformation for each principal quantum
number results in deformations for the intrinsic states of excited
baryonic states.  We find that the resulting baryon spectra agree very
well with the existing experimental data.  The spatial deformation of
the
baryonic excited states carry useful information on the quark
confinement
and provide a clue to understand the confining mechanism.  }

PACS:
\newpage

It is very important to understand the structure of hadrons for
the full understanding of the quark nuclear physics, which is governed
by
the QCD dynamics.
QCD is a non-abelian gauge
theory and exhibits many interesting non-perturbative phenomena.
Although quarks and gluons are the fundamental degrees
of freedom in the QCD
lagrangian, they are confined in hadrons.
At the same time, chiral
symmetry is spontaneously broken in the hadronic phase,
which provides the constituent masses to the quarks inside hadrons.
These non-perturbative properties in
turn play a crucial role in hadron structures and their interactions.

The direct theoretical study of QCD for the baryon spectra
including excited baryons in terms of lattice QCD seems difficult at
present and therefore we want to learn the physics of strong interaction
from the nature.
In this respect, there was an interesting suggestion
made a decade ago in the baryon spectra~\cite{TDD}.
The observation of the nucleon,
delta and lambda excitation spectra motivated a theoretical description,
where the ground state baryons such as the nucleon are spherical, but
the
excited baryons such as the Roper are deformed.
The use of the SU(3) version of the interacting boson model,
in which the Roper is identified
as the band head of the rotational band, provides spectra in good
agreement
with experiment.
The same observation was made by Bhaduri and his
collaborators~\cite{Bhaduri}.

These works were performed almost decade ago.  There is also an
elaborated model as the non-relativistic quark model with various
perturbative terms of Isgur and Karl~\cite{IK}.
The baryons as well as the meson
spectra are discussed also in terms of the group theoretical model by
Iachello and his collaborators~\cite{Iachello}.
Since the experimental facilities such as
CEBAF, the GeV photon facility at SPring-8 and the storage ring
LISS project at Indiana will provide variety of detailed data on the
hadron spectra, we believe it important to point out the interesting
regularity seen clearly in the baryon spectra~\cite{PDG}.
The theoretical view on
the non-perturbative phenomena are also advanced very much in these days
due to the detailed study of the dual Ginzburg-Landau (DGL)
theory~\cite{DGL} and
the lattice gauge theory on the confinement and the chiral symmetry
breaking mechanisms~\cite{KMS}.

Hence, the purpose of this paper is to introduce a model
for baryons with a spatially deformed shape
to describe the
confinement for quarks with the constituent quark mass due to chiral
symmetry breaking.
The spatial deformation is implemented by a simple deformed harmonic
oscillator potential, and therefore the model
is referred to as the deformed oscillator quark (DOQ) model.
The predictions of the
DOQ model is then compared with existing experimental data.

To begin with, we discuss a possible scenario of connection of the
DOQ model with QCD.
The QCD lagrangian consists of quarks and gluons,
which is written as~\cite{ChenLi}
\beq
\label{L_QCD}
L = \bar \psi \left( i \gamma_\mu \partial^\mu
+ e \gamma_\mu A^\mu - m \right) \psi
- \frac{1}{4} \tr G_{\mu \nu} G^{\mu \nu} \, .
\eeq
Here,  $\psi$ denotes
the quark field with current masses $m$  and $A^\mu$  are the gluon
fields.  QCD provides confinement of quarks and chiral symmetry breaking
as the most important non-perturbative phenomena for strong
interactions.
Although the explicit procedure is not yet known, we may construct the
baryonic states using the concept of the mean field approximation  on
gluon fields, which leads to the baryon lagrangian for quarks as
\beq
\label{QCD_mean}
L = \bar \psi \left( i \gamma_\mu \partial^\mu
-M(\bar A) - \phi (\bar A) \right) \psi
+ ({\rm higher\; order}) \, .
\eeq
Here, the masses $M(\bar A)$
of constituent quarks are acquired by spontaneous
chiral symmetry breaking.
The mean field, $\bar A$, should be provided by
knowing the quark configuration and its form may be highly complicated.
The confinement of quarks is expressed in terms of $\phi (\bar A)$,
which is considered
here as a scalar like function~\cite{Buchmueller}.
The constituent quark masses and the
confinement potential are functions of the mean fields of the gluon
fields, $\bar A$ , which are at the same time dependent on the quark
wave
function, $\psi$.
The higher order terms may contain, for example, Goldstone
boson exchange terms, perturbative gluon exchange terms and so on.
We may be able to formulate this scenario in terms of the dual
Ginzburg-Landau theory, which describes both confinement and chiral
symmetry breaking through the QCD monopole fields and their condensation
in the QCD vacuum~\cite{DGL}.

We take now the simplest possible form for the description of
baryons in terms of constituent quarks, which realizes the above
scenario.
We assume that the constituent quark masses are large enough
to take the non-relativistic quark model and the confinement potential
is expressed in terms of a deformed harmonic oscillator potential.
The hamiltonian is then written as
\beq
\label{hamiltonian}
H = \sum^3_{i=1} \left[ \frac{p_i^2}{2M_i} +
  \frac{1}{2} M_i ( \omega_x^2 x^2 + \omega_y^2 y^2
                  + \omega_z^2 z^2 )  \right]
\eeq
The oscillator parameters are assumed to satisfy
the volume conservation condition,
$\omega_x \omega_y \omega_z = \omega_0^3$  to minimize
the number of parameters.
We may relax this condition as our knowledge on
confinement becomes more clear.
Hence, there exist two free parameters in
the DOQ model, which are fixed by the nucleon and the Roper masses.  The
system of the deformed oscillator model is worked out for nuclei by Bohr
and Mottelson in detail and documented in their book~\cite{BM}.
The quarks acquire
the constituent masses due to chiral symmetry breaking, which
justifies the use of the non-relativistic form of the
kinetic energy.
For simplicity, we do not include more terms as the
pseudo-scalar degrees of freedom as naturally arisen as the Goldstone
bosons of chiral symmetry breaking~\cite{KMS}.
We solve three valence quark
systems with the removal of the center-of-mass coordinate, where the
intrinsic energy is expressed simply as
\beq
\label{Eint}
E_{\rm int} (N_x, N_y, N_z) =
\omega_x (N_x +1) + \omega_y (N_y +1) + \omega_z (N_z +1) \, ,
\eeq
where $N_i = n_i^\lambda +  n_i^\rho$
are the sum of the oscillator quanta of the intrinsic motion in
the Jaccobi coordinates, $\lambda$  and $\rho$.
We then take the variation of the
intrinsic energy (\ref{Eint}) with respect to deformation
$(\omega_x, \omega_y, \omega_z )$ for each principal
quantum number, $N = N_x + N_y + N_z$.
The results of this variation is
summarized in Table 1.
The lowest three states are $N = 0$, 1 and 2.
Their shapes are spherical for $N=0$, prolately
deformed with the ratio of 2 to 1
for $N=1$ and prolately deformed with the ratio of 3 to 1 for $N=2$.
It is then natural to identify the $N=0$ state with the nucleon,
the $N=1$ state with
the negative parity states around 1520 MeV and
the $N=2$ state with the
positive parity excited states.

We have to perform the angular momentum projection when the
intrinsic state is deformed.
This projection can be done by using the
Hill-Wheeler projection method~\cite{projection}.
The excitation energy of a state of orbital angular momentum $L$
is written as
\beq
\label{Erot}
E(N,L) = E_{\rm int}(N)
+ \frac{\hbar^2}{2I} \left[ L(L+1) - \bra L^2 \ket \right]
\eeq
$E_{\rm int}$ is obtained by the minimization condition for each $N$.
The moment of
inertia $I$ is calculated using the cranking formula~\cite{BM}.
$\bra L^2 \ket$ is the
average angular momentum of the intrinsic state.
These quantities are tabulated in Table 1.

Coupling the intrinsic spin $S$ of three quarks with the orbital
angular momentum $L$, take for example $S=1/2$, we construct
mass spectra for nucleon excited sates as shown in Fig. 1.
On the left hand side we show the theoretical results: one for the
positive parity states of $L=0, 2, 4, \cdots$ and the other
for the negative parity states of $L=1, 3, \cdots$.
In the theoretical side, two spin states $J = L \pm 1/2$
degenerate when
spin-orbit coupling is ignored, as experimental data suggest.
On the right hand side, experimental masses of well observed nucleon
excited states with four stars are shown [5].
One exception is the $5/2^{-}$ state of $D_{15}(2200)$ with two 
stars.  
This state is very likely to form a spin doublet with $G_{17}(2190)$.  
We do not list all the states but those which are well
identified with the $^2 8$ representations ($S = 1/2$)
of the spin-flavor group, and are well compared with the
DOQ model predictions.
The mass formula (5) contains two parameters; one is the
constant energy which determines the absolute value of
the ground state energy and the
other is the oscillator parameter $\omega_{0}$.
Hence Eq. (5) is essentially the formula with one parameter $\omega_0$.
The oscillator parameter $\omega_0$
is determined here by the average mass splitting
between the first excited states of $1/2^+$ (Roper like states)
and the corresponding ground states
for the flavor SU(3) baryons (for details of SU(3) baryons, see the
discussion below).
The resulting value is $\omega_0 = 607$ MeV.

Considering the simplicity of the DOQ model,
it is remarkable that many observed states fit very well
to the theoretical rotational band.
In particular, the first $1/2^{+}$ excited state, the Roper
resonance, appears near the first  $1/2^{-}$ excited states.
In the DOQ model, the energy subtraction due to the angular momentum
fluctuation,
$\hbar^{2} / 2I \langle L^{2} \rangle$,
is significant.
The sum of the rotational energy and the energy subtraction results in
the theoretical Roper state very close to the observed one.
If we look at the spectra in more detail,
theoretical masses of $1/2^-$ states appear lower than experiments.
Shifting up the theoretical masses by a constant amount,
the agreement would become better.

If the fundamental structure of the excitation spectrum is produced by
gluon dynamics through confinement and chiral symmetry breaking, we
should see a similar pattern in other members of the spin-flavor
group.  For this purpose, we tabulate in Table 2 well observed SU(3)
baryon excited states with three or four stars up to about 2 GeV.
In Figs. 2, the masses of the excited
states ($M^*$) measured from the corresponding ground state ($M$) are
shown as compared with theoretical predictions with
the oscillator parameter $\omega_0 = 607$ MeV as before.  Figs. 2a
and 2b are for positive and negative parity states.
There we have shown those states which seem to be well identified with
the rotational band of the DOQ model.
They are the representations which include the
ground state: $^{2}8$ for $N$, $\Lambda$ and $\Sigma$,  and
$^{4}10$ for $\Sigma$ and $\Delta$.  For
correspondence, we also show $^{2}10$ of negative parity
$\Delta$.
Here the notation is for the spin-flavor group, $^{2S+1} D$,
where $S$ is the total spin of the three quarks and $D$ the
representations of SU(3).
States are then specified by the coupling of $S$ and $L$.
Now we explain the identification of various states.

\noindent
{\it Ground states} \newline
        Because the three quarks are in the lowest S-state and form the
        symmetric spatial wave function, the possible spin-flavor states are
        $^{2}8$ and $^{4}10$.
        They form the $56^{+}$ dimensional representation.
        We do not discuss the mass differences within this multiplet,
        but rather the masses of the ground states are fixed as inputs on
        which excited states are constructed.
        Although multiplets such as $N(^{4}8)$, $\Lambda(^{2}1)$,
        $\Sigma(^{4}8)$, $\Delta(^{2}10)$ are not allowed for the ground
        states (this is shown in Table 2 by the symbol $--$), we keep those
        columns when the corresponding states exist in excited states.

\noindent
{\it Even parity excited states} \newline
        $N=2$ excited states yield the rotational
        band of $L = 0, 2, 4, \cdots$.
    Although, the spatial wave function can be
    either symmetric or mixed symmetric,
    we simply try first to put excited states into the spatially
    symmetric states.
    Then if there remain some, they will be identified with
    other allowed multiplets such as $^4 8$ ($S = 3/2$).
    It is interesting that many of the well observed excited states of
    the nucleon ($N$) fit into the rotational band of $^{2}8$.
    One exception is $P_{11}(1710)$ which would be identified with
    $^{4}8$.
    For the lambda ($\Lambda$) sector, again, more than half of the
    well observed states are identified with the $^{2}8$
    rotational band.
    The $^{4}8$ $\Lambda$ states have larger
   masses than those of the $^2 8$ multiplets.
   There should be spin correlations, which provide the difference
   between these multiplets.
    For $\Sigma$ states, in addition to the $^{2}8$ rotational band,
    there seems to be a strong evidence that
    there is a rotational band of the decouplet $^{4}10$.
    In order to emphasize this, we
    have also listed the one or two star states,
    $P_{13}(1840)$, $F_{15}(2070)$ and $P_{13}(2080)$.
    In particular, it is interesting that the one star state
    $P_{13}(1840)$ is likely to be identified with the Roper
    state of the decouplet $\Sigma$.
    In the delta ($\Delta$) sector, all the excited states up to $L=2$
    are observed and fit into the $^{4}10$ rotational band.

\noindent
{\it Odd parity excited states} \newline
        $N=1$ excited states yield the rotational
        band of $L = 1, 3, 5, \cdots$.
        For $L=1$, the spatial wave function is mixed symmetric because the
        symmetric one is for the spurious center of mass motion, and
        therefore, the allowed spin-flavor states are $^{2}8$, $^{4}8$,
        $^{2}1$ and $^{2}10$.
        They form the $70^-$ dimensional representation.
        For the nucleon sector, five excited states are clearly observed
        which may fit into $^{2}8$ and $^{4}8$ of $L=1$.
    As in the case of positive parity
    $\Lambda (^4 8)$, the masses of negative parity $N(^4 8)$ are
       larger than those of the $ ^2 8$ multiplets.
       Again the difference between these multiplets have to be reproduced
       with spin correlations.
        For the lambda sector,
        we have identified the observed seven states with $^{2}8$,
        $^{4}8$ and $^{2}1$.
        The $\Lambda(1405)$ and $\Lambda(1520)$ are identified with $^{2}1$,
        but are
        out of the systematics of the DOQ model, as they are much lighter
        than the model prediction.
        As discussed many times, these resonances, particularly
        $\Lambda(1405)$ would have a large $KN$ component
        rather than the single particle quark state~\cite{labda}.
        For the $\Sigma$ sector,
        we have identified three states with $^{2}8$ and
        $^{4}8$ of $L=1$.
        However, we can not exclude other possibilities, for instance,
        that they would belong to the same multiplet of $^{4}8$.
    For the delta sector, two states are identified with
    $^{2}10$ of $L=1$.
    There are several well known states for $\Xi$, but
    we do not list them here.
    There are three states that we do not list in the table (below 2
GeV):
    $\Delta\, S_{31}(1900)$, $\Delta\, D_{35}(1930)$ and
    $\Sigma\, D_{13}(1940)$.
    It seems to be difficult to put them in the systematics
    of the DOQ model.

In the present study of excited baryons,
we emphasize that most of the well observed states fit well to the
predictions of the DOQ model.  Furthermore,
the rotation like structure seems to be
rather universal in flavor SU(3), and therefore,
it is strongly implied that the dynamics of excited
baryons are dominantly flavor independent as provided by quark-gluon
interactions.

In conclusion, we have studied the baryon spectra using the
non-relativistic deformed oscillator quark model.
We find that excited states of various spin-flavor multiplet are
deformed.
The comparison of the spectrum of the DOQ model 
with experimental data is very good.  
In order to establish the underling mechanism of the baryon
spectra, we have to get more information experimentally as the higher
spin
states and gamma transitions.
Theoretically we have to calculate the
gamma transitions to distinguish various models, which are being
studied.

We are grateful to Prof. H. Ejiri for his fruitful and enlightning 
discussions on the present study.
We acknowledge also the collaboration of H. Fujimura at the early stage
of the present study.

\newpage

\subsection*{Table caption}

\begin{description}
        \item[Table 1:]  The properties of the deformed
        oscillator quark (DOQ) model for
each principal quantum number $N$.
$E_{\rm int}$ denotes the intrinsic energy in
unit of $\hbar \omega_0$
with the ratio of the oscillator parameter,
$\omega_x : \omega_y : \omega_z$,
which provides the shape of the intrinsic state.
When the intrinsic state is deformed as
indicated in the column of shape,
the rotational states appear with the
moment of inertia, $\hbar^2 /2I$.
$\bra L^2 \ket$  denotes the average value of the
orbital angular momentum for
each deformed intrinsic state.

        \item[Table 2:]  Baryon states of $uds$ flavors are
categorized to the predictions of the DOQ model.
  All the details of the identification of the experimentally
  observed states in this table are described in the text.

\end{description}

\subsection*{Figure caption}

\begin{description}
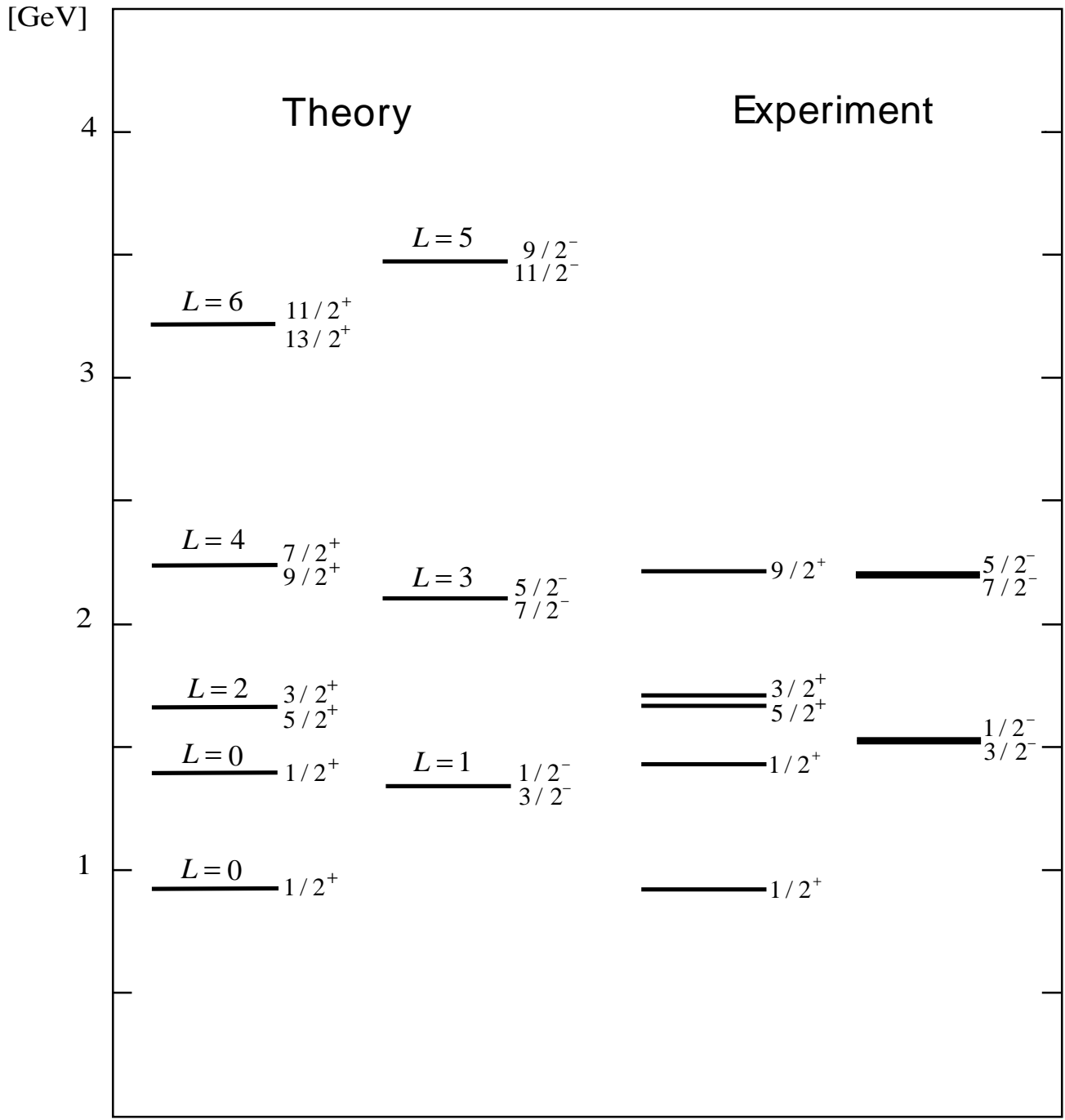
        \item[Figure 1:]  The energy spectrum of nucleon resonances for
positive and
negative parity states in unit of GeV.  In the theoretical
spectrum, $L$ for each state denotes the orbital angular momentum.
The intrinsic spin, $S=1/2$, is coupled to $L$
to form total spin states,
which are denoted in the right hand side of each state.
The rotational bands are formed on the first excited positive
and the negative parity states.
The experimental spectrum is shown
in the right hand side of this figure.

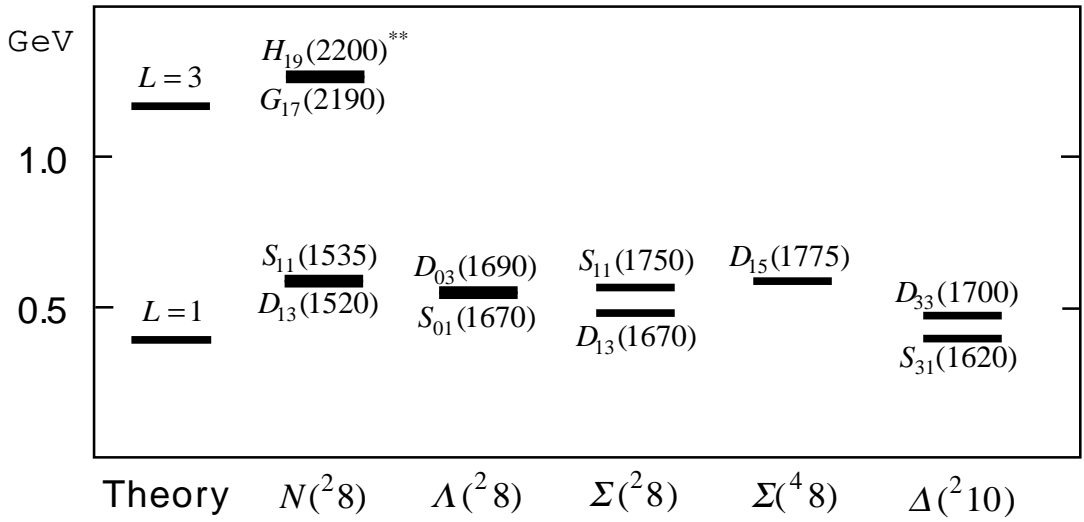
        \item[Figure 2:]  The mass spectrum of the positive and the negative
parity
        states for $uds$ baryons in unit of GeV.
        The masses are measured from the ground states of each spin-flavor
    quantum number.
    The theoretical prediction is shown in the left part
    for each parity.
\end{description}

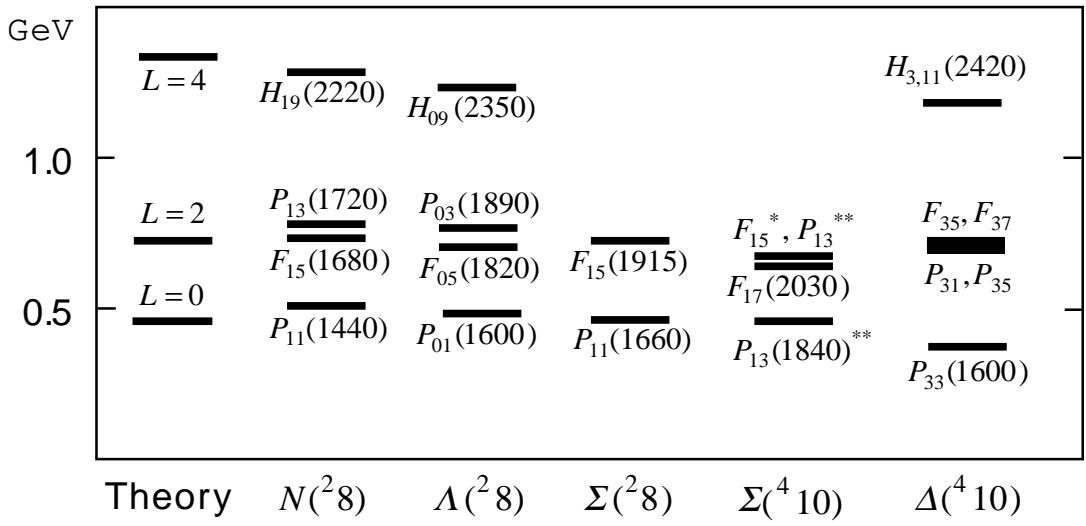
\end{document}